\documentclass[pra,twocolumn,groupedaddress,superscriptaddress,amsmath,amssymb,a4paper]{revtex4}
\usepackage{geometry}
\usepackage{graphicx}
\usepackage{mathrsfs}
\usepackage{amssymb}
\usepackage{amsmath}
\usepackage{amsthm}
\usepackage{amsbsy}
\usepackage{bm}
\usepackage{hyperref}
\usepackage[T1]{fontenc}
\usepackage{color}

\theoremstyle{definition}

\newcommand{\bra}[1]{\langle #1|}
\newcommand{\ket}[1]{| #1 \rangle }
\newcommand{\ip}[2]{{\langle #1|}{ #2 \rangle }}

\begin{document}

\title{Quantum dynamics intervened by repeated nonselective measurements}

\author{Sergey N. Filippov}

\affiliation{Moscow Institute of Physics and Technology,
Institutskii Per. 9, Dolgoprudny, Moscow Region 141700, Russia}

\affiliation{Institute of Physics and Technology of the Russian
Academy of Sciences, Nakhimovskii Pr. 34, Moscow 117218, Russia}

\begin{abstract}
We derive the theory of open quantum system dynamics intervened by
a series of nonselective measurements. We analyze the cases of
time independent and time dependent Hamiltonian dynamics in
between the measurements and find the approximate master equation
in the stroboscopic limit. We also consider a situation, in which
the measurement basis changes in time, and illustrate it by
nonselective measurements in the basis of diabatic states of the
Landau-Zener model.
\end{abstract}

\maketitle

\section{Introduction}

Quantum measurements are extensively used in practice to get the
information about the system in interest (estimation~\cite{paris},
discrimination~\cite{chefles},
comparison~\cite{filippov-ziman-2012},
tomography~\cite{buzek-1998}), to modify the quantum state
(conditional state preparation~\cite{resch}, quantum
control~\cite{pechen-2006,shuang-2008,pechen-2015}), and to probe
parameters of physical fields (quantum sensing~\cite{degen}).
Quantum states are described by density operators that are
positive semi-definite and have unit trace. In the process of a
selective measurement, observation of outcome $x$ results in
transformation of the initial density operator $\varrho$ into the
conditional output state of the form
\begin{equation}
\label{conditional-output-state} \varrho \rightarrow \varrho_x =
\frac{{\cal I}_x [\varrho] }{ {\rm tr}\big[ {\cal I}_x [\varrho]
\big]},
\end{equation}

\noindent where ${\cal I}_x$ is a completely positive
trace-decreasing map (quantum operation)~\cite{hz}. Physically,
$p_x = {\rm tr} \left[ \mathcal{I}_x [\varrho] \right]$ is a
probability of observing outcome $x$. The mapping $x \rightarrow
{\cal I}_x$ is usually referred to as an instrument. Disregarding
measurement outcomes, we get a so-called nonselective measurement:
\begin{equation}
\varrho \rightarrow \Phi[\varrho] = \sum_{x} p_x \varrho_x =
\sum_{x} \mathcal{I}_x [\varrho].
\end{equation}

\noindent Physical requirement $\sum_x p_x = 1$ implies that ${\rm
tr} \left[ \sum_x \mathcal{I}_x [\varrho] \right] = {\rm
tr}[\varrho]$ for all density operators $\varrho$, i.e. $\Phi$ is
trace preserving. $\Phi$ is also completely positive as a sum of
completely positive maps, so $\Phi$ is nothing else but a quantum
channel~\cite{holevo-book}. Thus, a general nonselective
measurement is merely a quantum channel. On the other hand, any
quantum channel $\Phi$ can be represented in the operator-sum form
$\Phi[\varrho] = \sum_{m} A_m \varrho A_m^{\dag}$, where $\sum_{m}
A_m^{\dag} A_m = I$ (identity operator). Therefore, a quantum
channel $\Phi$ can be treated as a nonselective measurement with
instrument operations $\mathcal{I}_m [\varrho] = A_m \varrho
A_m^{\dag}$.

If $\mathcal{I}_x[\varrho] = {\rm tr}[E_x \varrho] \omega_x$,
where $x \rightarrow E_x$ is a positive operator-valued measure
(POVM) with POVM elements $E_x$, $\{\omega_x\}$ is a set of
density operators, then such a nonselective measurement describes
a so-called measure-and-prepare procedure (a channel of Holevo
form)~\cite{hz,holevo-book}. The corresponding channel $\Phi$ is
entanglement breaking in this
case~\cite{holevo-book,horodecki-2003}.

Suppose that $\{\ket{k}\}_{k=1}^d$ is an orthonormal basis in
$d$-dimensional Hilbert space ${\cal H}_d$ and $\varrho$ is a
density operator on ${\cal H}_d$. Consider a measurement with $d$
outcomes and corresponding instrument $\mathcal{I}_k[\varrho] =
\bra{k} \varrho \ket{k} \ket{k}\bra{k}$. Nonselective realization
of such a measurement leads to the following transformation of the
density operator:
\begin{equation}
\label{nonselective-map} \varrho \rightarrow \Lambda[\varrho] =
\sum_{k=1}^d \bra{k} \varrho \ket{k} \ket{k}\bra{k}.
\end{equation}

\noindent Physically, such a process describes the complete
dephasing in the basis $\{\ket{k}\}_{k=1}^d$. Denote
$\varrho_{kk'} = \bra{k} \varrho \ket{k'}$ matrix elements of the
density operator $\varrho$, then $\bra{k} \Lambda [\varrho]
\ket{k} = \varrho_{kk}$ and $\bra{k} \Lambda [\varrho] \ket{k'} =
0$ if $k \neq k'$. As a result, quantum coherence in the state
$\Lambda[\varrho]$ is completely destroyed. It is
Eq.~\eqref{nonselective-map}, which is usually referred to as
nonselective measurement in the
literature~\cite{pechen-2006,shuang-2008,pechen-2015}. In what
follows, we also focus on nonselective measurement channels
$\Lambda$ of the form \eqref{nonselective-map}.

Clearly, two nonselective measurements $\Lambda$ performed one
after another do not change the system state, i.e.
\begin{equation}
\label{Lambda-Lambda} \Lambda \circ \Lambda = \Lambda.
\end{equation}

If the duration between nonselective measurements is not equal to
zero, then the system state evolves in time. Although nondiagonal
elements $\varrho_{kk'}$ vanish after each nonselective
measurement, diagonal elements $\varrho_{kk}$ have enough time to
change. Repeated nonselective measurements can be used in quantum
control of system evolution. For instance, appropriately choosing
the time moments of nonselective measurements one can accelerate
the probability of Landau-Zener transitions~\cite{pechen-2015}.
Apparently, if measurements are performed with high repetition
rate, the system becomes frozen (in analogy with the conventional
Zeno effect~\cite{misra-sudarshan-1977}).

The goal of this paper is to derive an approximate master
equation, which describes the system evolution intervened by a
large number of repeated nonselective measurements. Such a master
equation describes the evolution of the diagonal density operator,
which effectively simulates the actual density operator at the
moments just after nonselective measurements.

The paper is organized as follows. In
Sec.~\ref{section-master-equation}, we derive a stroboscopic limit
of quantum evolution and the governing master equation for a
general case of nonselective measurements. In
Sec.~\ref{section-different-bases}, we consider the situation,
where the measurement basis changes in time. In
Sec.~\ref{section-LZ}, we apply the derived theory to Landau-Zener
transitions intervened by nonselective measurements. In
Sec.~\ref{section-conclusions}, brief conclusions are given.

\section{Master equation for quantum dynamics with repeated nonselective measurements}
\label{section-master-equation}

In the absence of measurements, a general quantum open system
dynamics is described by a convolutionless master equation
\begin{equation}
\label{convolutionless} \frac{d\varrho(t)}{dt} = L(t)
[\varrho(t)],
\end{equation}

\noindent where $L(t)$ is a time-dependent
generator~\cite{breuer-petruccione-2002}. A formal solution of
this equation is $\varrho(t) = T_{\leftarrow} \exp \left(
\int_{-\infty}^t L(t') dt' \right) \varrho(-\infty)$, where
$T_{\leftarrow}$ is the Dyson time-ordering operator. In the case
of unitary evolution with the Hamiltonian $H(t)$, $L(t)[\cdot] =
-i [H(t),\cdot]$. Hereafter we assume the Planck constant to be
equal to 1.

To get the characteristic timescale $\frac{1}{\gamma(t)}$ of
changes in the state $\varrho(t)$, we extract the characteristic
strength of the generator $L(t)$ by formula
\begin{equation}
\label{gamma} L(t) = \gamma(t) {\cal L}(t),
\end{equation}

\noindent where the map ${\cal L}(t)$ is dimensionless and its
Schatten norm $\|{\cal L}(t)\|_{1 \rightarrow 1} = 1$.

Suppose that the evolution~\eqref{convolutionless} is intervened
by nonselective measurements~\eqref{nonselective-map}, which are
performed at sequential time moments $t_1, \ldots, t_N$. In the
time interval $(t_n,t_{n+1})$ the exact dynamics with such
nonselective measurements defines the dynamical map
\begin{eqnarray}
\label{Lambda-exact-arbitrary} \Phi(t) &=& T_{\leftarrow} \exp
\left( \int_{t_n}^t L(t') dt' \right) \circ \Lambda \circ \ldots
\nonumber\\
&& \circ \Lambda \circ T_{\leftarrow} \exp \left(
\int_{-\infty}^{t_1} L(t') dt' \right),
\end{eqnarray}

\noindent $\varrho(t) = \Phi(t)[\varrho(-\infty)]$. We assume that
the initial state $\varrho(-\infty)$ is an eigenstate of
$\Lambda$, i.e. $\Lambda[\varrho(-\infty)] = \varrho(-\infty)$
(otherwise, one can count time from the instance of the first
measurement), so one can formally perform the nonselective
measurement at time $t_0 = -\infty$. At time moments $t=t_n$ the
transformation $\Phi_{t}$ maps any operator into ${\rm
supp}\Lambda$, so the output is diagonal in the basis
$\{\ket{k}\}_{k=1}^d$. Since we are interested in the derivation
of master equation for the diagonal density operator, we consider
only time moments $t_n$ and consider the map $\widetilde{\Phi}(t)$
by $\Lambda \Phi(t) \Lambda$. Using property \eqref{Lambda-Lambda}
and formula \eqref{gamma}, we get
\begin{equation}
\widetilde{\Phi}(t_n) = \prod_{m=1}^{n} \Lambda \circ
T_{\leftarrow} \exp \left( \int_{t_{m-1}}^{t_m} \gamma(t') {\cal
L}(t') dt' \right) \circ \Lambda.
\end{equation}

We assume that the nonselective measurements are performed in such
a way that $\int_{t_{m-1}}^{t_m} \gamma(t') dt' \ll 1$. In this
case the map $T_{\leftarrow} \exp \left( \int_{t_{m-1}}^{t_m}
L(t') dt' \right)$ is close to the identity transformation ${\rm
Id}$ and $\| T_{\leftarrow} \exp \left( \int_{t_{m-1}}^{t_m} L(t')
dt' \right) - {\rm Id}\|_{1 \rightarrow 1} \ll 1$. Roughly
speaking, the system cannot evolve too far from the diagonal state
between the measurements. Such a regime corresponds to a
sufficiently high repetition rate of measurements ($\sim |t_m -
t_{m-1}|^{-1}$) as compared to the characteristic frequency of
system evolution $\gamma(t)$. If this is the case, then with the
accuracy up to the second order of $\int_{t_{m-1}}^{t_m}
\gamma(t') dt'$ we have
\begin{eqnarray}
&& \Lambda \circ T_{\leftarrow} \exp \left( \int_{t_{m-1}}^{t_m}
\gamma(t')
{\cal L}(t') dt' \right) \circ \Lambda \nonumber\\
&& \approx \Lambda + \int_{t_{m-1}}^{t_m} dt' \gamma(t') \Lambda
{\cal L}(t') \Lambda \nonumber\\
&& \quad + \frac{1}{2} \int_{t_{m-1}}^{t_m} dt'
\int_{t_{m-1}}^{t_m} dt'' \gamma(t') \gamma(t'') \Lambda
T_{\leftarrow} {\cal L}(t') {\cal
L}(t'') \Lambda \nonumber\\
&& \approx \Lambda \circ T_{\leftarrow} \exp \Bigg[
\int_{t_{m-1}}^{t_m} dt'
\gamma(t') \Lambda {\cal L}(t') \Lambda \nonumber\\
&& \qquad + \frac{1}{2} \int_{t_{m-1}}^{t_m} dt'
\int_{t_{m-1}}^{t_m} dt'' \gamma(t') \gamma(t'') \nonumber\\
&& \qquad\quad \times \Big( \Lambda {\cal L}(t') {\cal L}(t'')
\Lambda - \Lambda {\cal L}(t') \Lambda {\cal L}(t'')
\Lambda \Big) \Bigg] \circ \Lambda \nonumber\\
&& = \Lambda \circ T_{\leftarrow} \exp \left[ \int_{t_{m-1}}^{t_m}
dt' L_{\rm eff}(t') \right] \circ \Lambda,
\end{eqnarray}

\noindent where the effective generator $L_{\rm eff}(t')$
preserves the diagonal structure of the density operator and reads
\begin{eqnarray}
\label{L-eff} && \!\!\!\!\!\!\!\!\!\! L_{\rm eff}(t') = \gamma(t')
\Lambda {\cal L}(t')
\Lambda \nonumber\\
&& \!\!\!\!\! + \frac{1}{2} \int_{t_{m-1}}^{t_m} dt'' \gamma(t')
\gamma(t'') \Big( \Lambda {\cal L}(t') {\cal L}(t'') \Lambda -
\Lambda {\cal L}(t')
\Lambda {\cal L}(t'') \Lambda \Big) \nonumber\\
&& \!\!\!\!\!\!\!\!\!\!  = \Lambda L(t') \Lambda + \frac{1}{2}
\int_{t_{m-1}}^{t_m} dt'' \Big( \Lambda L(t') L(t'') \Lambda -
\Lambda L(t') \Lambda L(t'')
\Lambda \Big) \nonumber\\
&& \\
&& \!\!\!\!\!\!\!\!\!\!  = \Lambda L(t') \left[ {\rm Id} +
\frac{1}{2} \int_{t_{m-1}}^{t_m} dt'' \Big( L(t'') - \Lambda
L(t'') \Lambda \Big) \right] \Lambda. \nonumber
\end{eqnarray}

Therefore, the transformation $\varrho(t_{m-1}) \rightarrow
\varrho(t_{m})$ of the density operator $\varrho(t)$ from one time
moment $t_{m-1}$ of nonselective measurement to the next time
moment $t_m$ of nonselective measurement is approximately given by
the master equation
\begin{equation}
\frac{d\varrho(t)}{dt} = L_{\rm eff}(t) [\varrho(t)].
\end{equation}

If the measurements are performed stroboscopically after equal
time periods $\tau$ and the generator $L = \gamma {\cal L}$ is
time independent, then we obtain the effective dynamical semigroup
generator
\begin{equation}
\label{effective-generator} L_{\rm eff} = \gamma \Lambda {\cal L}
\Lambda + \frac{\gamma^2 \tau}{2} \Big( \Lambda {\cal L}^2 \Lambda
- \Lambda {\cal L} \Lambda {\cal L} \Lambda \Big).
\end{equation}

This form of the generator naturally appears in the so-called
stroboscopic limit $\gamma\tau \rightarrow 0$ and $\gamma^2 \tau
\rightarrow {\rm const} \neq 0$~\cite{luchnikov-2017}.

The obtained form of the effective generator becomes particularly
easy in the case of Hamiltonian dynamics, when $L[\cdot] = \gamma
{\cal L}[\cdot] = -i \gamma [h, \cdot]$. Using notation $h_{kk'}$
for matrix elements $\bra{k} h \ket{k'}$, we get
\begin{eqnarray}
&& \!\!\!\!\!\!\!\!\!\!  \Lambda {\cal L} \Lambda[X] = -i \sum_{k,l=1}^d \bra{k} X \ket{k} \,\, \bra{l} \left[ h, \ket{k}\bra{k} \right] \ket{l} \,\, \ket{l}\bra{l} = 0, \label{Lambda-L-Lambda} \\
&& \!\!\!\!\!\!\!\!\!\!  \Lambda {\cal L} \Lambda {\cal L} \Lambda[X] = \Lambda {\cal L} \Big[ \Lambda {\cal L} \Lambda[X] \Big] = 0, \label{Lambda-L-Lambda-L-Lambda} \\
&& \!\!\!\!\!\!\!\!\!\!  \Lambda {\cal L}^2 \Lambda [X] = -\sum_{k
\neq k'} |h_{kk'}|^2 \Big( \{ \ket{k}\bra{k}, X \} - 2
\ket{k'}\bra{k} X \ket{k} \bra{k'} \Big) \nonumber\\
\end{eqnarray}

\noindent and find out that the effective dynamical semigroup
generator \eqref{effective-generator} has the celebrated
Gorini-Kossakowski-Sudarshan-Lindblad
form~\cite{gks-1976,lindblad-1976}:
\begin{equation}
\label{L-eff-constant-Hamiltonian} L_{\rm eff} [X] \!=\! \gamma^2
\tau \!\sum_{k \ne k'}\! |h_{kk'}|^2 \!\left(\! \ket{k'}\bra{k} X
\ket{k} \bra{k'} \!-\! \frac{1}{2} \big\{ \ket{k}\bra{k}, X \big\}
\!\right)\!.
\end{equation}

The transition operators $h_{k'k}\ket{k'}\bra{k}$, $k \neq k'$ are
the Lindblad operators, which are responsible for the
redistribution of the diagonal elements of the density operator.
The evolution of diagonal elements $\varrho_{kk}$ of the density
operator has the simple form of the classical Pauli equation:
\begin{eqnarray}
\label{Pauli-eq} \frac{\partial \varrho_{kk}(t)}{\partial t} =
\sum_{k' \ne k} \big( W_{k' \rightarrow k} \, \varrho_{k'k'}(t) -
W_{k \rightarrow k'} \, \varrho_{kk}(t) \big),
\end{eqnarray}

\noindent where $W_{k' \rightarrow k} = \gamma^2 \tau |h_{kk'}|^2$
and $\sum_{k' \ne k} W_{k \rightarrow k'} = \gamma^2 \tau \left(
\bra{k} h^2 \ket{k} - \bra{k} h \ket{k}^2 \right)$.
Eq.~\eqref{Pauli-eq} shows that the quantum system under repeated
nonselective measurements with $\gamma\tau \rightarrow 0$,
$\gamma^2\tau \rightarrow {\rm const}$ evolves like a classical
statistical system. Noting that $W_{k' \rightarrow k} = W_{k
\rightarrow k'}$, we get
\begin{eqnarray}
\label{Pauli-eq-simple} \frac{\partial \varrho_{kk}(t)}{\partial
t} = \sum_{k' \ne k} W_{k' \rightarrow k} \big( \varrho_{k'k'}(t)
- \varrho_{kk}(t) \big),
\end{eqnarray}

\noindent so the fixed point of the derived dynamical map is the
maximally mixed state $\varrho = \frac{1}{d} I$.

Returning to the discussion of general formula~\eqref{L-eff}, the
second term in the right hand side of Eq.~\eqref{L-eff} is of the
order $\gamma(t') \int_{t_{m-1}}^{t_m} \gamma(t'') dt'' \ll
\gamma(t')$. If one neglects this term and considers the
Hamiltonian dynamics $L(t)[\cdot] = \gamma(t) {\cal L}(t)[\cdot] =
-i \gamma(t) [h(t), \cdot]$, then $L_{\rm eff}(t) = \Lambda L(t)
\Lambda = 0$ and the usual Zeno effect is
reproduced~\cite{misra-sudarshan-1977}, i.e. the dynamics of the
density operator becomes effectively frozen. Taking into account
the second term in the right hand side of Eq.~\eqref{L-eff}, one
can see the deviation from the trivial dynamics, with the
characteristic timescale of the change in the density operator
being $\left( \gamma(t') \int_{t_{m-1}}^{t_m} \gamma(t'') dt''
\right)^{-1} \gg \frac{1}{\gamma(t')}$.

In the case of purely dissipative dynamics $L(t)[X] = \gamma(t)
{\cal L}(t)[X]$, ${\cal L}(t)[X] =  A(t) X A^{\dag}(t) -
\frac{1}{2} \{ A^{\dag}(t)A(t),X \}$, the first term $\Lambda L(t)
\Lambda$ in Eq.~\eqref{L-eff} already gives nonzero contribution
to $L_{\rm eff}(t)$. Neglecting the second term in
Eq.~\eqref{L-eff}, we get
\begin{eqnarray}
\frac{\partial \varrho_{kk}(t)}{\partial t} = \gamma(t) &
\displaystyle\sum\limits_{k' \neq k} & \bigg(
|\bra{k}A(t)\ket{k'}|^2
\varrho_{k'k'}(t) \nonumber\\
&& - |\bra{k'}A(t)\ket{k}|^2 \varrho_{kk}(t) \bigg).
\end{eqnarray}

\noindent Thus, the dissipative dynamics intervened by
nonselective measurements also results in the classical Pauli
equation for diagonal elements of the density matrix.

\section{Repeated nonselective measurements in different bases}
\label{section-different-bases}

Suppose that nonselective measurements at time moments
$t_1,\ldots,t_N$ are performed in different bases, then the
corresponding map $\Lambda$ is time dependent, i.e. nonselective
measurement at time moment $t_n$ results in a map $\Lambda_n$.
Continuing the same line of reasoning as in the previous section,
one can readily obtain the transformation $\varrho(t_{m-1})
\rightarrow \varrho(t_m)$ and the effective generator describing
it:
\begin{eqnarray}
&& \Lambda_m \circ T_{\leftarrow} \exp \left( \int_{t_{m-1}}^{t_m}
L(t') dt' \right) \circ \Lambda_{m-1} \nonumber\\
&& = \Lambda_m \Lambda_{m-1} +
T_{\leftarrow} \exp \left( \int_{t_{m-1}}^{t_m} L_{\rm eff}(t') dt' \right) - {\rm Id}, \quad \label{transformation-total-bases} \\
&& L_{\rm eff}(t') = \Lambda_m L(t') \Lambda_{m-1}  \nonumber\\
&& \qquad\qquad + \frac{1}{2} \int_{t_{m-1}}^{t_m} dt'' \Big(
\Lambda_m L(t')
L(t'') \Lambda_{m-1} \nonumber\\
&& \qquad\qquad\quad - \Lambda_m L(t') \Lambda_{m-1} \Lambda_m
L(t'') \Lambda_{m-1} \Big). \label{L-eff-different-measurements}
\end{eqnarray}

Since measurement bases change in time, one should separately
treat the term $\Lambda_{m} \Lambda_{m-1} [X] =
\sum_{k_{m},k_{m-1}} |\ip{k_{m}}{k_{m-1}}|^2
\ket{k_{m}}\bra{k_{m-1}} X \ket{k_{m-1}}\bra{k_{m}}$. The map
$\Lambda_{m} \Lambda_{m-1}$ describes transition from a matrix
diagonal in basis $\{k_{m-1}\}$ to a matrix diagonal in basis
$\{k_{m}\}$. In what follows, we will focus on transformations of
diagonal elements of the density matrix $\{\varrho_{kk}\}$, where
the basis changes in time accordingly. In such an approach, the
map $\Lambda_{m-1} \Lambda_m$ acts as follows:
\begin{equation}
\varrho' = \Lambda_{m-1} \Lambda_m [\varrho] \Longleftrightarrow
\varrho_{k'k'}' = \sum_{k} |\ip{k_{m}'}{k_{m-1}}|^2 \varrho_{kk},
\end{equation}

\noindent where the latter formula describes the transformation of
diagonal elements. Each matrix $B^{(m,m-1)}$ with elements
$B_{k'k}^{(m,m-1)} = |\ip{k_{m}'}{k_{m-1}}|^2$ is doubly
stochastic and can be represented in the form
\begin{equation}
B^{(m,m-1)} = \exp\left( Q^{(m,m-1)} (t_m - t_{m-1}) \right),
\end{equation}

\noindent where $Q^{(m,m-1)}$ is the transition rate matrix
satisfying conditions $Q_{kk}^{(m,m-1)} = - \sum_{k' \neq k}
Q_{k'k}^{(m,m-1)} = - \sum_{k' \neq k} Q_{kk'}^{(m,m-1)}$ and
$Q_{k'k}^{(m,m-1)} \geqslant 0$ if $k \neq k'$. Physically, the
matrix $Q$ shows how quickly the measurement basis
$\{\ket{k(t)}\}$ changes in time, namely, $Q_{k'k}(t) \approx
\left\vert \frac{\partial \bra{k'(t)}}{\partial t} \ket{k(t)}
\right\vert^2 (t_m - t_{m-1})$ if $k \neq k'$ and the measurements
are performed quite often that linear approximation $\ket{k'(t_m)}
\approx \ket{k'(t_{m-1})} + (t_m - t_{m-1}) \left. \frac{\partial
\ket{k'(t)}}{\partial t} \right\vert_{t = t_{m-1}}$ is valid.

Similarly, the transformation $\varrho' = L_{\rm eff} [\varrho]$
with $L_{\rm eff}$ in the
form~\eqref{L-eff-different-measurements} can be equivalently
described by such a matrix $R^{(m,m-1)}$ that $\varrho_{k'k'}' =
\sum_{k} R_{k'k}^{(m,m-1)} \varrho_{kk}$. Rewriting
Eq.~\eqref{transformation-total-bases} in the differential form
for diagonal elements of the density operator, we get
\begin{equation}
\frac{\partial \varrho_{kk}(t)}{\partial t} = \sum_{l} \Big(
Q_{kl}(t) + R_{kl}(t) \Big) \varrho_{ll}(t). \label{Q-R}
\end{equation}

The approximate form of matrix $R^{(m,m-1)}$ can be found in the
case of unitary evolution with a time dependent Hamiltonian
$H(t)$, when $L[\cdot] = -i [H(t),\cdot]$, such that vectors
$\{\ket{k_{m-1}}\}$ are eigenvectors of $H(t_{m-1})$ and vectors
$\{\ket{k_{m}}\}$ are eigenvectors of $H(t_{m})$. Then $\Lambda_m
L(t_{m-1}) \Lambda_{m-1} \approx 0$ and $\Lambda_m L(t_m)
\Lambda_{m-1} \approx 0$, so for $t \in (t_{m-1},t_m)$ one can
assume that $\Lambda_m L(t') \Lambda_{m-1} \approx 0$. Similarly,
$\Lambda_m L(t') \Lambda_{m-1} \Lambda_m L(t'') \Lambda_{m-1}
\approx 0$, and Eq.~\eqref{L-eff-different-measurements} reduces
to
\begin{eqnarray}
L_{\rm eff}(t') & \approx & \frac{1}{2} \int_{t_{m-1}}^{t_m} dt''
\Lambda_m L(t') L(t'') \Lambda_{m-1} \nonumber\\
& \approx & \frac{1}{2} \int_{t_{m-1}}^{t_m} dt'' \Lambda_m
L^2(t'') \Lambda_{m-1}.
\end{eqnarray}

Using formula~\eqref{Pauli-eq}, we see that the effective
Lindbladian induces evolution of diagonal elements in the form
$\frac{\partial \varrho_{kk}(t)}{\partial t} = \sum_{l \neq k}
R_{kl}(t) \Big( \varrho_{ll}(t) - \varrho_{kk}(t) \Big)$, where
$R_{kl}(t) = \int_{t_{m-1}}^{t_m} |\bra{k_m} H(t'')
\ket{l_{m-1}}|^2 dt''$, $l \neq k$. Exploiting the approximation
$\ket{k(t_m)} \approx \ket{k(t_{m-1})} + (t_m - t_{m-1}) \left.
\frac{\partial \ket{k(t)}}{\partial t} \right\vert_{t = t_{m-1}}$,
we evaluate
\begin{equation}
R_{kl}(t) \approx \left\vert \frac{\partial \bra{k(t)}}{\partial
t} \ket{l(t)} \right\vert^2 E_l^2(t) (t_m - t_{m-1})^3,
\end{equation}

\noindent where $E_l(t)$ is the eigenvalue of $H(t)$ corresponding
to eigenvector $\ket{l(t)}$. Combining results of this section, at
time interval $t \in (t_{m-1},t_m)$ Eq.~\eqref{Q-R} reduces to
\begin{eqnarray}
\frac{\partial \varrho_{kk}(t)}{\partial t} & = & \sum_{l \neq k}
\left\vert \frac{\partial \bra{k(t)}}{\partial t} \ket{l(t)}
\right\vert^2 \Big( 1 + E_l^2(t) (t_m - t_{m-1})^2 \Big) \nonumber\\
&& \times (t_m - t_{m-1}) \Big( \varrho_{ll}(t) - \varrho_{kk}(t)
\Big), \label{Q-R-simple}
\end{eqnarray}

\noindent provided the measurements are performed often enough
that the linear approximation $\ket{k'(t_m)} \approx
\ket{k'(t_{m-1})} + (t_m - t_{m-1}) \left. \frac{\partial
\ket{k'(t)}}{\partial t} \right\vert_{t = t_{m-1}}$ remains valid.

\section{Landau-Zener transitions with nonselective measurements}
\label{section-LZ}

Consider a two-level system with the orthonormal basis
$\{\ket{0},\ket{1}\}$. Let the system occupy the state $\ket{0}$
at time $t=-\infty$, i.e. $\varrho(-\infty) = \ket{0}\bra{0}$. The
time-dependent Hamiltonian of Landau-Zener model reads
\begin{equation}
H(t) = \Delta \sigma_x + \epsilon t \sigma_z,
\end{equation}

\noindent where $\Delta,\epsilon > 0$, $\sigma_x = \ket{0}\bra{1}
+ \ket{1}\bra{0}$, and $\sigma_z = \ket{0}\bra{0} -
\ket{1}\bra{1}$. Solution of the equation $\frac{d\varrho(t)}{dt}
= - i[H(t),\varrho(t)]$ can be found analytically~\cite{brataas}.
Equation
\begin{equation}
\bra{0} \varrho(+\infty) \ket{0} = \exp \left( - \frac{\pi
\Delta^2}{\epsilon} \right)
\end{equation}

\noindent represents the Landau-Zener formula for transition
probability from the lower energy level at $t=-\infty$ (state
$\ket{0}$) to the higher energy level at $t=+\infty$ (state
$\ket{0}$)~\cite{landau,zener}. If the rate $\epsilon \rightarrow
\infty$, then $\bra{0} \varrho(+\infty) \ket{0} \rightarrow 1$,
which describes the situation of a sudden change of the
Hamiltonian, when the system state does not have enough time to
evolve. If the rate $\epsilon \rightarrow 0$, then $\bra{0}
\varrho(+\infty) \ket{0} \rightarrow 0$, which describes the
adiabatic regime with constant population of the ground state.

Suppose the process of the evolution is intervened by repeated
nonselective measurements in the basis of instantaneous
eigenvectors of $H(t)$ (also referred to as diabatic states):
\begin{eqnarray}
&& \ket{\varphi_0(t)} = \frac{ \left( \sqrt{\Delta^2 + (\epsilon
t)^2} - \epsilon t \right) \ket{0} - \Delta \ket{1}
}{\sqrt{2\sqrt{\Delta^2 + (\epsilon t)^2} \left(\sqrt{\Delta^2 +
(\epsilon t)^2} - \epsilon t \right)}}, \quad \\
&& \ket{\varphi_1(t)} = \frac{ \Delta \ket{0} + \left(
\sqrt{\Delta^2 + (\epsilon t)^2} - \epsilon t \right) \ket{1}
}{\sqrt{2\sqrt{\Delta^2 + (\epsilon t)^2} \left(\sqrt{\Delta^2 +
(\epsilon t)^2} - \epsilon t \right)}}. \quad
\end{eqnarray}

More precisely, at time moment $t_m$ one performs the nonselective
measurement in the basis $\ket{0_m} = \ket{\varphi_0(t_m)}$,
$\ket{1_m} = \ket{\varphi_1(t_m)}$. Our goal is to minimize the
transition rate via selective measurements, i.e. to preserve
system in the ground state.

Using Eq.~\eqref{Q-R-simple}, we get the effective dynamics of
population of diabatic level $1$ between $(m-1)$-th and $m$-th
measurements:
\begin{eqnarray}
\frac{\partial \varrho_{11}(t)}{\partial t} &=& \left\vert
\frac{\partial \bra{\varphi_1(t)}}{\partial t} \ket{\varphi_0(t)}
\right\vert^2 \Big( 1 + E_0^2(t) (t_m - t_{m-1})^2 \Big)
\nonumber\\
&& \times (t_m - t_{m-1}) \Big( \varrho_{00}(t) -
\varrho_{11}(t) \Big). \label{LZ-1}
\end{eqnarray}

Direct calculation yields $E_0^2(t) = \Delta^2 + (\epsilon t)^2$
and
\begin{equation}
\left\vert \frac{\partial \bra{\varphi_1(t)}}{\partial t}
\ket{\varphi_0(t)} \right\vert^2 = \frac{\epsilon^2
\Delta^2}{4\Big(\Delta^2 + (\epsilon t)^2 \Big)^2}.
\end{equation}

\noindent Taking into account that $\varrho_{00}(t) = 1 -
\varrho_{11}(t)$, we get
\begin{eqnarray}
\label{p11} && \frac{\partial \varrho_{11}(t)}{\partial t} =
\frac{\epsilon^2 \Delta^2 (t_m - t_{m-1}) }{4\left(\Delta^2 +
(\epsilon t)^2 \right)} \nonumber\\
&& \times \left( \frac{1}{\Delta^2 + (\epsilon t)^2} + (t_m -
t_{m-1})^2 \right) \Big( 1 - 2 \varrho_{11}(t) \Big).\quad
\end{eqnarray}

The obtained equation is valid if $t_m - t_{m-1} \ll
\left(\Delta^2 + (\epsilon t)^2 \right) / \epsilon \Delta$.
Periods between selective measurements can be rather long if $t
\rightarrow - \infty$ or $t \rightarrow + \infty$. Therefore, the
moments of nonselective measurements can be chosen within the
range $t_m,t_{m-1}\in[-2\Delta/\epsilon,2\Delta/\epsilon]$.
Basically, the application of nonselective measurements beyond
that interval would not substantially affect the population
dynamics.

Consider the regime $\epsilon \gg \Delta^2$, when the probability
to remain in the ground state $\varrho_{00}(+\infty)$ tends to 0,
whereas the transition probability $\varrho_{11}(+\infty)$ tends
to 1. Let us demonstrate that the transition probability can be
effectively suppressed by a finite number of nonselective
measurements. For instance, distribute $2N+1$ measurements
uniformly within the interval
$[-2\Delta/\epsilon,2\Delta/\epsilon]$, so $t_m - t_{m-1} =
\frac{4\Delta}{\epsilon 2N}$, then integration of Eq.~\eqref{p11}
yields
\begin{eqnarray}
\varrho_{11}(+\infty) & = & \frac{1}{2}\left[ 1 - \exp\left( -
\frac{\pi (8 \Delta^4 + \epsilon^2 N^2)}{2 \epsilon^2 N^3} \right)
\right] \nonumber\\
& \approx & \frac{1}{2} \left[ 1 - \exp\left( - \frac{\pi}{2N}
\right) \right] \approx \frac{\pi}{4N},
\end{eqnarray}

\noindent where approximations are made under condition $N \gg 1$.

Alternatively, since the transition rate dominates around $t=0$,
one can can distribute instances of nonselective measurements more
often in the vicinity of $t=0$, for instance, $t_m = 2 \Delta
(m-N-1)(m-N) / \epsilon N (N+1)$, $m=1, \ldots, 2N+1$. This
distribution corresponds to substitution $t_m - t_{m-1} = 2
\Delta^2 t / (N+1) \left( \Delta^2 + (\epsilon t)^2 \right)$ in
Eq.~\eqref{p11}, which results in
\begin{eqnarray}
\varrho_{11}(+\infty) & = & \frac{1}{2}\left[ 1 - \exp\left( -
\frac{(N+1)^2 + \frac{4 \Delta^4}{3 \epsilon^2}}{2 (N+1)^3}
\right) \right] \nonumber\\
& \approx & \frac{1}{2} \left[ 1 - \exp\left( - \frac{1}{2N}
\right) \right] \approx \frac{1}{4N}.
\end{eqnarray}

This means that one can preserve the population of the ground
state not only in the case of slow rate $\epsilon \ll \Delta^2$,
but also for a fast rate $\epsilon \gg \Delta^2$ with the
appropriate use of repeated nonselective measurements.

With the use of one nonselective measurement the transition rate
can be diminished at least to one half~\cite{pechen-2015}. We
conjecture that with the use of $N$ nonselective measurements the
minimal transition probability in the regime $\epsilon \gg
\Delta^2$ diminishes as $O(\frac{1}{N})$.

\section{Conclusions}
\label{section-conclusions}

We have analyzed the quantum dynamics intervened by repeated
projective rank-1 nonselective measurements. Such measurements
make the density operator diagonal in the measurement basis.
However, due to a possibility to evolve in between the
measurements, quantum system does not remain frozen in general. We
have derived the approximate master equation describing the
evolution of diagonal elements of the density operator and
demonstrated its relation with the classical Pauli equation in the
so-called stroboscopic limit.

We have studied the case of repeated nonselective measurements,
when measurement bases change with time. In the case of the
unitary dynamics, when the measurement basis changes in time in
accordance with the eigenvectors of instantaneous Hamiltonian, we
find a simplified evolution equation. The developed theory is
applied to the Landau-Zener model. We show that appropriate
nonselective measurements can reduce the transition probability
between the ground and excited state even in the case of the fast
rate of Hamiltonian.

\bigskip

\begin{acknowledgments}

The author is grateful to the anonymous referee for many valuable
comments to improve the quality of the manuscript and, in
particular, for suggestions of simpler forms of
Eqs.~\eqref{Lambda-L-Lambda}-\eqref{Lambda-L-Lambda-L-Lambda}. The
study is supported by the Russian Foundation for Basic Research
under Project No. 16-37-60070 mol-a-dk.

\end{acknowledgments}

\end{document}